\pdfoutput=1
\RequirePackage[T1]{fontenc}
\PassOptionsToPackage{T1}{fontenc}
\PassOptionsToPackage{full}{textcomp}
\PassOptionsToPackage{pdftex,
pdfencoding=auto,
pdfnewwindow=true,
pdfusetitle=true,
bookmarks=true,
bookmarksnumbered=true,
bookmarksopen=true,
pdfpagemode=UseThumbs,
bookmarksopenlevel=1,
pdfpagelabels=false,
breaklinks=true,
colorlinks=true
}{hyperref}
\PassOptionsToPackage{usenames,dvipsnames,table}{xcolor}
\PassOptionsToPackage{square,sort&compress,comma,numbers,elide}{natbib}
\PassOptionsToPackage{utf8}{inputenc}
\documentclass[twocolumn,
10pt,
aps,
prx,
english,
superscriptaddress,
twoside,
floatfix,
nofootinbib,
notitlepage,
longbibliography]{revtex4-2}
\usepackage[T1]{fontenc}
\setcitestyle{square,sort&compress,comma,numbers,elide}

\makeatletter
\def\bstctlcite{\@ifnextchar[{\@bstctlcite}{\@bstctlcite[@auxout]}}
\def\bstctlcite[#1]#2{\@bsphack
\@for\@citeb:=#2\do{%
\edef\@citeb{\expandafter\@firstofone\@citeb}%
\if@filesw\immediate\write\csname #1\endcsname{\string\citation{\@citeb}}\fi}%
\@esphack}
\makeatother

\usepackage{mathtools}

\usepackage[utf8]{inputenx}
\usepackage[english]{babel}
\usepackage[style=american,autopunct=true]{csquotes}
\usepackage{units}
\usepackage{lmodern}
\usepackage{microtype}
\usepackage{soul}
\usepackage{centernot}

\usepackage{appendix}

\usepackage{graphicx}
\usepackage{float}
\usepackage[usenames,dvipsnames,table]{xcolor}
\definecolor{purple}{RGB}{128,0,128}
\definecolor{ultramarine}{RGB}{63, 0, 255}
\definecolor{medblue}{RGB}{0, 0, 100}
\definecolor{panblue}{RGB}{0,24,150}
\definecolor{carmine}{RGB}{150, 0, 24}
\definecolor{gray}{RGB}{150, 150, 150}
\definecolor{googleblue}{RGB}{34, 0, 204}
\definecolor{darkgreen}{RGB}{0, 80, 0}

\usepackage{paralist}
\usepackage{hyperref}
\hypersetup{colorlinks,
linkcolor=carmine,
citecolor=darkgreen,
urlcolor=googleblue,
anchorcolor=OliveGreen}

\usepackage{amssymb,amsthm,amsfonts,amstext,amsmath}
\DeclareMathAlphabet\mathbfcal{OMS}{cmsy}{b}{n}

\usepackage{url}

\def\be{\begin{equation}}
\def\ee{\end{equation}}
\def\bea{\begin{eqnarray}}
\def\eea{\end{eqnarray}}
\def\bma{\begin{mathletters}}
\def\ema{\end{mathletters}}

\def\q0{\underline{0}}

\def\G{{\cal G}}

\def\id{{\mathbb I}}

\def\one{\leavevmode\hbox{\small1\normalsize\kern-.33em1}}

\newtheorem{theo}{Theorem}

\newtheorem{prob}{Problem}

\newtheorem*{openq}{Open Question}
\def\id{{\mathbb I}}
\usepackage{dsfont}
\def\ds1{{\mathds{1}}}

\definecolor{darkBluishGreen}{RGB}{0,79,160}
\newcommand{\term}[1]{\textcolor{medblue}{\textbf{\upshape #1}}}

\makeatletter
\renewcommand{\@seccntformat}[1]{\csname the#1\endcsname\quad}
\renewcommand*{\thesection}{\arabic{section}}
\renewcommand*{\thesubsection}{\thesection.\arabic{subsection}}
\renewcommand*{\p@subsection}{}
\renewcommand*{\thesubsubsection}{\thesubsection.\arabic{subsubsection}}
\renewcommand*{\p@subsubsection}{}
\makeatother
\usepackage{secdot}
\sectionpunct{section}{\quad} 
\sectionpunct{subsection}{\quad} 
\newcommand{\nocontentsline}[3]{}
\newcommand{\tocless}[2]{\bgroup\let\addcontentsline=\nocontentsline#1{#2}\egroup}

\begin{document}

\interfootnotelinepenalty=10000
\title{The Inflation Technique Completely Solves the Causal Compatibility Problem}
\author{Miguel Navascu\'es}
\affiliation{Institute for Quantum Optics and Quantum Information (IQOQI) Vienna, \\\hspace{1em}Austrian Academy of Sciences, Boltzmanngasse 3, 1090 Vienna, Austria}
\author{Elie Wolfe}
\affiliation{Perimeter Institute for Theoretical Physics, \\\hspace{1em}31 Caroline St. N, Waterloo, Ontario, Canada, N2L 2Y5}

\begin{abstract}
The causal compatibility question asks whether a given causal structure graph --- possibly involving latent variables --- constitutes a genuinely plausible causal explanation for a given probability distribution over the graph's observed categorical variables. Algorithms predicated on merely \emph{necessary} constraints for causal compatibility typically suffer from false negatives, i.e. they admit incompatible distributions as \emph{apparently} compatible with the given graph. In \href[pdfnewwindow]{https://doi.org/10.1515/jci-2017-0020}{DOI:\textbf{10.1515/jci-2017-0020}}, one of us introduced the \emph{inflation technique} for formulating useful relaxations of the causal compatibility problem in terms of linear programming. In this work, we develop a formal hierarchy of such causal compatibility relaxations. We prove that inflation is asymptotically tight, i.e., that the hierarchy converges to a zero-error test for causal compatibility. In this sense, the inflation technique fulfills a longstanding desideratum in the field of causal inference. We quantify the rate of convergence by showing that any distribution which passes the $n^{th}$-order inflation test must be ${O}{\left(n^{\nicefrac{{-}{1}}{2}}\right)}$-close in Euclidean norm to some distribution genuinely compatible with the given causal structure. Furthermore, we show that for many causal structures, the (unrelaxed) causal compatibility problem is faithfully formulated already by either the first or second order inflation test.
\end{abstract}

\maketitle

\section{Introduction}
A Bayesian network or \term{causal structure} is a directed acyclic graph (DAG) where vertices represent random variables, each of which is generated by a non-deterministic function depending on the values of its parents. Nowadays, causal structures are commonly used in bioinformatics, medicine, image processing, sports betting, risk analysis, and experiments of quantum nonlocality. In this work we consider causal structures with two distinct types of vertices:  categorical variables which may be directly observed, and variables which cannot be observed, referred to as latent variables.\footnote{\citet{Pearl2009} refers to such graphical models as ``latent structures".} We make no assumption whatsoever on the state spaces of the latent variables; they can be discrete or continuous. Nevertheless, every causal structure encodes a possible \emph{hypothesis} of causal explanation for statistics over its observed variables. 

Naturally, understanding how different causal structures give rise to different sets of compatible distributions is a fundamental goal within the field of causal inference. Many prior works are ultimately concerned with the \term{causal discovery problem}, which asks to enumerate (or graphically characterize) all legitimate hypotheses of causal structure which are capable of explaining some observed probability distribution~\cite{Chickering2002b,Koller2009,Spirtes2016,TatradReview,Wermuth2015,Lauritzen2018,Magliacane2016,Evans2018ModelSelection}.  
For computational tractability, practical causal discovery algorithms typically exclude causal explanations which are unfaithful (fine-tuned). Fundamentally, however, the faithfulness assumption is not an essential criterion for causal discovery. Demanding faithfulness can be thought of as a \emph{second} filtering step, where the \emph{fundamental} filtering of causal discovery is the exclusion of any causal structure which cannot explain the observed probability distribution, even granting fine-tuning. In this manuscript, therefore, causal discovery refers to the foundational problem of returning \emph{all} causal structures compatible with the given distribution.  Selecting a single \enquote{best} causal model --- or even \emph{scoring} the quality of the different causal explanations~\cite{Magliacane2016,Evans2018ModelSelection} --- constitute refinements to the causal discovery problem which we do not address here.

The \term{causal characterization problem} is the focus of a distinct line of research. It concerns characterizing the set of statistics compatible with a single given causal structure, that is, the derivation of causal compatibility constraints~\cite{extraRef0,extraRef1,extraRef2,extraRef3,extraRef4,
SteudelAy,evans2012graphicalmethods,Richardson2013SingleWI,Evans2018NMP,Evans2018ModelSelection}. Quantum information theorists have recently joined this research effort~\cite{fritz2012bell,
chaves2014informationinference,RossetNetworks,tavakoli2016noncyclic,
ChavesPolynomial,Miklin2017Entropic,Kela2017Covariance,Weilenmann2017}. Causal characterization is useful for proving the impossibility of simulating certain quantum optics experiments with classical devices~\cite{BilocalCorrelations,Tavakoli2017BellToNetwork,Chaves2017starnetworks}, or for confirming the nonclassicality of quantumly realizable statistics in novel hypothetical scenarios~\cite{Wood2015,HLP,Pienaar2017,Himbeeck2018instrumental,Fraser2018triangle}.

We must note that the causal characterization problem has also been tackled in scenarios where the state-spaces of the latent variables are prescribed \cite{knownLatent1,knownLatent2,knownLatent3,moreGeigerMeek}. Critically, Ref.~\cite{rosset2016finite} provides upper bounds on the cardinalities of a causal structure's latent variables without any loss of generality (whenever the observed variables have discrete state spaces). Consequently, the set of multi-variate categorical distributions compatible with any given causal structure is always a semi-algebraic set, admitting characterization in terms of a finite number of polynomial equality and inequality constraints. Nevertheless, \emph{identifying} the full set of causal compatibility constraints via exploiting the constrained state-spaces of the latent variables is often intractable~\cite{knownLatent1,knownLatent2,knownLatent3,moreGeigerMeek}. The inflation technique considered herein, by contrast, has no dependence on the latent variables' state-spaces. Hereafter, therefore, we consider only causal structures with \emph{unconstrained} latent variables.


Causal discovery relates a single distribution to many structures;
 causal characterization relates many distributions to a single structure. Both such efforts, therefore, are oracle-wise equivalent, and hinge fundamentally on the \term{causal compatibility problem (CCP)}, which simply asks a yes-or-no question: Is the given distribution compatible with the given causal structure?
The \term{inflation technique} \cite{wolfe2016inflation} is a way of relating approximations of the causal compatibility problem to linear programming (LP) problems. Every LP \emph{satisfaction} problem can be dualized and recast as an equivalent \emph{optimization} problem. Inspired by LP duality, we will formulate a dual notion of causal compatibility, through which we will be able to rigorously upper-bound the error introduced by approximating the CCP as an LP problem via inflation. Our main result here is that this error asymptotically tends to zero when inflation is expressed as a hierarchy of ever-higher-order tests of causal compatibility. This implies that the inflation criterion --- far from being a relaxation --- is meta-equivalent to the causal compatibility problem, and hence constitutes an alternative way of understanding general causal structures.

In contrast with Ref.~\cite{wolfe2016inflation}, in this paper we define the inflation technique as a hierarchy of causal compatibility tests applicable \emph{exclusively} to the special class of causal structures (introduced by \citet{fritz2012bell}) called \term{correlation scenarios}. At the same time, however, we also introduce a \emph{graphical preprocessing} which precisely recasts the general causal compatibility problem in terms of causal compatibility with correlation scenarios, such that there is no loss of generality in our approach.

This paper is organized as follows: In Section~\ref{definitions} we introduce the concept of a correlation scenario, we define primal and dual notions of the causal compatibility problem and their approximations. In Section~\ref{overview} we review the inflation technique as a means for approximately solving either form of the causal compatibility problem. 
In Section~\ref{convergencenoproof} we state our main theorems concerning the convergence of inflation for correlation scenarios, though the formal proofs are deferred to the appendices. 
In Section~\ref{extension} we build upon existing causal inference techniques to describe a natural graphical preprocessing which maps general causal structures into correlation scenarios. This preprocessing --- fairly useful in its own right --- implies the universal applicability of the inflation technique as defined here. 
Finally, Section~\ref{conclusions} presents our conclusions.

\begin{figure}
  \centering
    \includegraphics[scale=0.8]{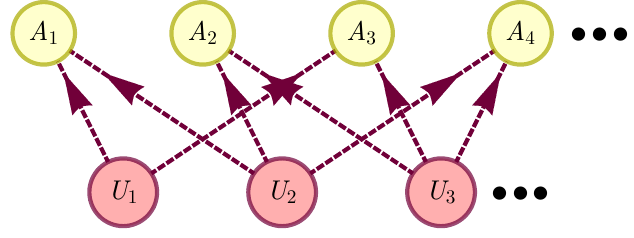}
  \caption{\textbf{A generic correlation scenario.} The independent latent variables $U_1,U_2{,}...$ influence the observed variables $A_1,A_2{,}...$.}
  \label{causal_network}
\end{figure}

\section{Preliminary Definitions}
\label{definitions}

The graphical models we study are fully general causal structures. A causal structure is represented by a directed acyclic graph imbued with some distinction among the vertices to clarify if a node in the graph represents either an observable or a latent variable. In this work, we use a pink color and subscripts of the letter \enquote{$U$} (\enquote{$U$} from ``Unobserved'') to indicate the latent variables in a graph. We follow the convention of Refs.~\cite{extraRef4,Shpitser2014NMP,Evans2018NMP} and depict exogenous (i.e., non-root) observable variables as square-shaped nodes in their graphs. 

\begin{figure}[b]
  \centering
  \includegraphics[scale=0.8]{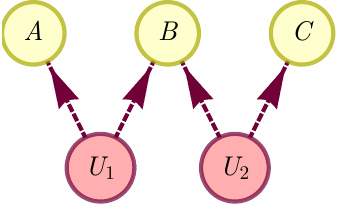}
  \caption{\textbf{The {three-on-line} correlation scenario.}}
  \label{bilocality}
\end{figure}

Correlation scenarios are a special type of causal structures. The graph of a correlation scenario has just two layers: a bottom layer of independently distributed latent random variables $\{U_1,U_2{,}...,U_L\}$ and a top layer of observable random variables $\{A_{1},A_2{,}...,A_m\}$, see Fig.~\ref{causal_network}. The observable distribution $P(A_{1},A_2{,}...,A_m)$ is generated via non-deterministic functions $A_x=A_{x}(\bar{U}^{L_x})$, with $\bar{U}=(U_1{,}...,U_L)$ and $L_x\subset\{1{,}...,L\}$. Here (and in the following) the notation $\bar{v}^S$, where $\bar{v}$ is a vector with $N$ entries and $S\subset \{1{,}...,N\}$, will represent the vector with entries $v^s$, $s\in S$. Readers familiar with \term{${\boldsymbol{d}\textbf{-separation}}$} may appreciate that although the implication $(A_i \perp_d A_j | A_k)\implies (A_i \perp_d A_j A_k)$ does \emph{not hold} for general causal structures, it \emph{is true} for correlation scenarios. Later on in Section~\ref{extension}, we will relate distributions over general causal structures to distributions over some correlation scenarios associated with them. As we will see, correlation scenarios are the atomic constituents upon which the inflation hierarchy acts.

\subsection{The Causal Compatibility Problem, its dual and their approximate versions}

A distribution $P$ over the observable variables of a causal structure $\G$ is said to be \term{compatible} with $\G$ if $P$ is the observable \emph{marginal} of some distribution $P'$ over \emph{all} the variables of $\G$, and where $P'$ can be factored into a product of singleton-variable conditional probability distributions associated with every individual vertex in $\G$ (conditioned on all of the vertex's parents, if any). A diverse vocabulary of phrases synonymous with \enquote{$P$ is compatible with $\G$} can be found in conventional literature, such as \enquote{$P$ can be \emph{realized} in $\G$}, \enquote{$P$ can \emph{arise} from $\G$}, \enquote{$\G$ \emph{gives rise} to $P$}, \enquote{$\G$ \emph{explains} $P$}, \enquote{$\G$ can \emph{simulate} $P$}, and \enquote{$\G$ is a \emph{model} for $P$}. 

\begin{figure}
  \centering
  \includegraphics[scale=0.8]{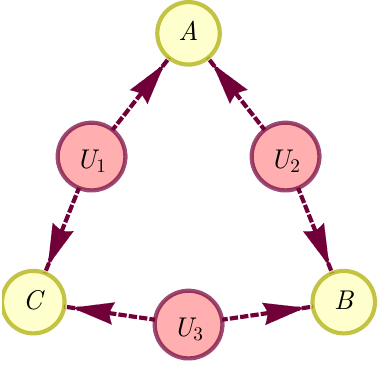}
  \caption{\textbf{The triangle scenario.}}
  \label{triangle^pic}
\end{figure}

Consider, for example, the correlation scenario dubbed \term{the triangle scenario}, with $m=L=3$, see Fig.~\ref{triangle^pic}. Denoting $A_1, A_2,A_3$ respectively by $A,B,C$, we have that a probability distribution $P(A,B,C)$ is realizable in the triangle scenario if $A,B,C$ are generated via the non-deterministic functions $A(U_1,U_2), B(U_2,U_3), C(U_3,U_1)$. Alternatively, $P(A,B,C)$ is realizable in the triangle scenario if it admits a decomposition of the form:
\begin{align}\label{realization}\begin{split}
&P(A,B,C)=
\\
&\smashoperator[l]{\sum_{U_1,U_2,U_3}}
\binom{P(A|U_1,U_2)P(B|U_2,U_3)P(C|U_3,U_1)}{
\times P(U_1)P(U_2)P(U_3)}.\end{split}
\end{align}

\begin{figure}
  \centering
  \includegraphics[scale=0.8]{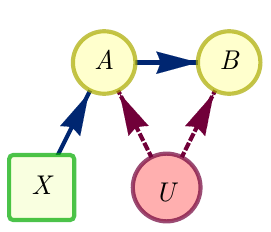}
  \caption{\vtop{\hbox{\textbf{The instrumental scenario.}}(Not a correlation scenario.)}}
  \label{instrumental}
\end{figure}

For causal structures which are \emph{not} correlation scenarios, however, the non-deterministic functions giving rise to the observed variables will also depend on other observed variables. An example is given by the \term{instrumental scenario}~\cite{Pearl95Instrumental,Himbeeck2018instrumental} (Fig.~\ref{instrumental}, up), where $X$ and $U$ are, respectively, a free observable and a latent variable, and the observed variables $A$ and $B$ are generated via the non-deterministic functions $A=A(X,U),B=B(A,U)$. Alternatively, $P(A,B|X)$ is realizable in the instrumental scenario if it admits a decomposition of the form:
\begin{align}
P(&A,B|X)
=\smashoperator[l]{\sum_{U}}
P(A|X,U)P(B|A,U)P(U).
\end{align}

Per Ref.~\cite{rosset2016finite}, the set of distributions $P$ compatible with a given causal structure $\G$ is a semi-algebraic set whenever the observable random variables are categorical, i.e., when they have finite cardinality. This implies that it can be characterized in terms of a finite number of polynomial inequalities. Unfortunately, the computational complexity of deriving such a characterizing set of inequalities makes the problem intractable already for networks of a very small size \cite{algorithms}. Furthermore, within the context of quantum foundations, there exist fairly natural causal structures for which the total number of such inequalities grows exponentially with the dimensionality of $P$ \cite{werner}. We must resort thus to partial characterizations of the original set of distributions. This notion is better formalized by the following problem.

\begin{prob}{\term{Approximate Causal Compatibility}}\label{causal^def}\\
\noindent\textsc{Input:} $\epsilon>0$, a causal structure $\G$ and a particular probability distribution $P$ over the observed variables.\\
\noindent\textsc{Output:} If there does not exist a probability distribution $\tilde{P}$ over the observed variables, such that $\|P-\tilde{P}\|^2\leq\epsilon$ and $\tilde{P}$ is compatible with $\G$, then return a function $F$ such that $F(P)<0$ and $F(\hat{P})\geq 0$ for all distributions $\hat{P}$ compatible with $\G$.\\
\noindent\textsc{Objective:} Determine if $P$ is ``approximately compatible'' with $\G$; if not, provide a witness $F$ to prove incompatibility of $P$.
\end{prob}

\noindent Note that, if $P$ is not approximately compatible with $\G$, the function $F$ witnessing its incompatibility is not required to be \emph{universal}. Namely, there could exist other distributions $P'$ incompatible with $\G$ such that $F(P')\geq 0$.

The goal of this paper is to provide a solution to this problem. Note that, since for any $\G$ the set of compatible distributions is closed \cite{rosset2016finite}, it follows that any distribution $P$ that is $\epsilon$-compatible for all $\epsilon>0$ must be compatible with $\G$. The analog problem for $\epsilon=0$ will be simply referred to as \emph{Causal Compatibility}.

A related problem that we will also solve is the following:

\begin{prob}{\term{Approximate Causal Optimization}}\label{App^Opt}\\
\noindent\textsc{Input:} $\epsilon>0$, a causal structure $\G$ and a real function $F$ of the probabilities of the observed events.\\
\noindent\textsc{Output:} A real value $f$ such that ${f\leq f_\star=\smashoperator{\min_{\hat{P}}}F(\hat{P})\leq f+\epsilon}$, where the minimum is over all distributions $\hat{P}$ compatible with $\G$.\\
\noindent\textsc{Objective:} Given a function $F$, find a good lower bound on its minimum value over all distributions compatible with $\G$.
\end{prob}

This problem is dual to Approximate Causal Compatibility, and it is interesting in its own right. In quantum optics experiments, we test and quantify non-classicality via the violation of inequalities of the form $F(P)\geq K$. Identifying values of $K$ for which the above holds for all distributions $P$ compatible with the considered causal structure $\G$ is a must before any experiment is actually carried out. Similarly as before, for $\epsilon=0$, we name the analog problem \emph{Causal Optimization}.

Coming back to the triangle scenario, an instance of Approximate Causal Optimization would be minimizing
\begin{equation}
(P(0,0,0)-1/2)^2+(P(1,1,1)-1/2)^2
\label{eq:GHZrobustness}
\end{equation}
over all distributions $P(A,B,C)$ with $A,B,C\in\{0,1\}$ compatible with the triangle scenario. In any experimental setup where bipartite optical sources play the role of $U_1,U_2,U_3$ in the triangle scenario, any observed distribution $P(A,B,C)$ for which the value of (\ref{eq:GHZrobustness}) is smaller than the lower bound $f$ provided by Approximate Causal Optimization evidences the presence of quantum effects.

There exist a number of algorithms which provide outer approximations for the set of distributions compatible with a given causal structure $\G$ \cite{extraRef4,Kela2017Covariance,Weilenmann2017,Koller2009, Shpitser2014NMP, outer1, outer2}. By minimizing functions over such outer approximations, existing algorithms can provide lower bounds on the true minimum and thus solve Approximate Causal Optimization, as long as $\epsilon$ exceeds some threshold (determined by the mismatch between the aforementioned relaxations and the original set of compatible distributions). As we will see, the Inflation Technique can be used to solve both Approximate causal compatibility and Approximate Causal Optimization for arbitrarily small values of $\epsilon$.

\section{The Inflation Hierarchy for Correlation Scenarios}
\label{overview}

\subsection{Some examples}

\begin{figure}
  \centering
    \includegraphics[scale=0.8]{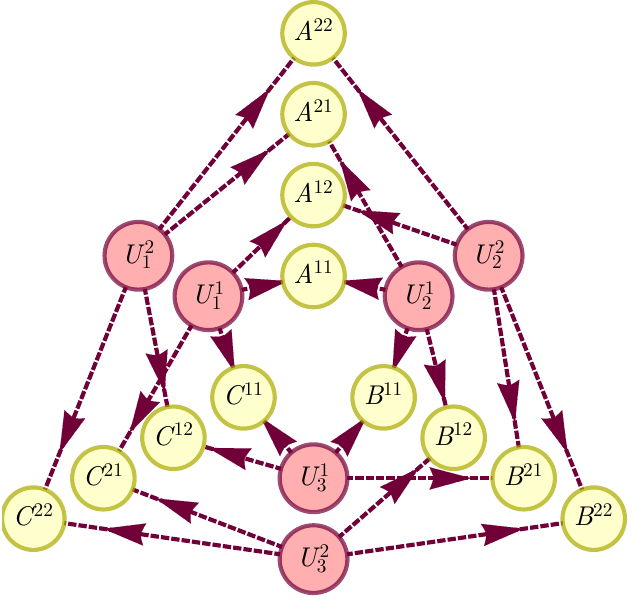}
  \caption{\textbf{The second-order inflation graph of the triangle scenario.}}
  \label{inflation^pic}
\end{figure}

Let $P(A,B,C)$ be a distribution realizable in the triangle scenario, and suppose that we generate $n$ independently distributed copies of $U_{1},U_2,U_3$, that is, the variables $\{U^i_1,U^i_2,U^i_3:i=1{,}...,n\}$. Then we could define the random variables 
\begin{align}\label{eq:tridependencies}
\hspace{-1.75ex}A^{ij}\equiv A(U^i_1,U^j_2),B^{ij}\equiv B(U^i_2,U^j_3),C^{ij}\equiv C(U^i_3,U^j_1).
\end{align}
\noindent  The causal structure associated with the independently distributed copies of $U_{1},U_2,U_3$ and their observable children $\left\{\{A^{ij}\},\{B^{kl}\},\{C^{pq}\}\right\}$ is termed an \term{inflation graph}. The inflation graph of a correlation scenario is also a correlation scenario; as an example, Fig.~\ref{inflation^pic} depicts the ${n=2}$ inflation graph for the triangle scenario. These observable variables follow a probability distribution $Q_n(\{A^{ij}\},\{B^{kl}\},\{C^{pq}\})$ with the property
\begin{align}\label{symmetry}
&Q_n(\{A^{ij}{=}a^{ij}, B^{kl}{=}b^{kl}, C^{pq}{=}c^{pq}\})=\\\nonumber
&Q_n(\{A^{ij}{=}a^{\pi(i)\pi'(j)},B^{kl}{=}b^{\pi'(k)\pi''(l)}, C^{pq}{=}c^{\pi''(p)\pi(q)}\}),
\end{align}
\noindent for all permutations of $n$ elements $\pi,\pi',\pi''$. Expanded out for ${n=2}$, Eq.~\eqref{symmetry} becomes 
\begin{align*} Q_2&(a^{11},a^{12},a^{21},a^{22},b^{11},b^{12},b^{21},b^{22},c^{11},c^{12},c^{21},c^{22})\\\nonumber
= &Q_2(a^{21},a^{22},a^{11},a^{12},b^{11},b^{12},b^{21},b^{22},c^{12},c^{11},c^{22},c^{21})
\\\nonumber
= &Q_2(a^{12},a^{11},a^{22},a^{21},b^{21},b^{22},b^{11},b^{12},c^{11},c^{12},c^{21},c^{22})
\\\nonumber
= &Q_2(a^{11},a^{12},a^{21},a^{22},b^{12},b^{11},b^{22},b^{21},c^{21},c^{22},c^{11},c^{12})
\end{align*}

We treat with special distinction the \term{diagonal variables} $\{A^{ii},B^{ii},C^{ii}\}_i$. Given the global distribution $Q_n$, we denote by $Q_n^g$ the marginal distribution of the diagonal variables with indices up to $g\leq n$, i.e.
\begin{align}\label{diagtriangle}
Q_n^g\coloneqq\;Q_n\big(\bigwedge\limits_{i=1}^g \{A^{ii}{=}a^i,B^{ii}{=}b^i,C^{ii}{=}c^i\}\big)
\end{align}
\noindent In the following, we call $Q_n^g$ the \term{diagonal marginal of degree-\boldmath$g$}.

A related concept is the \term{degree-\boldmath$g$ lifting} of a distribution $P$, consisting of the statistics of $g$ independent and identically distributed copies of $P$, that is
\begin{align*}
P^{\otimes g}\big(\bigwedge\limits_{i=1}^g \{A^{ii}{=}a^i,B^{ii}{=}b^i,C^{ii}{=}c^i\}\big)\coloneqq \;\prod_{i=1}^g P(a^i,b^i,c^i).
\end{align*}
Taking the random variables in the inflation graph to arise per Eq.~\eqref{eq:tridependencies} implies that the diagonal marginals associated with the inflation graph must be related to the lifted distributions of the original distribution over observed variables per
\begin{align}
& Q_n^g = P^{\otimes g}, \mbox{ for }g=1{,}...,n.
\label{probas}
\end{align}
Expanded out for ${g=n=2}$, Eq.~\eqref{probas} identifies the diagonal marginal in this scenario ${Q_{n{=}2}^{g{=}2}={\sum_{A^{12},A^{21},B^{12},B^{21},C^{12},C^{21}}}
Q_{n{=}2}}$ as 
\begin{align}\begin{split}
Q_{n{=}2}^{g{=}2}&(a^{11},a^{22},b^{11},b^{22},c^{11},c^{22})\\
&=P(a^{11},b^{11},c^{11})P(a^{22},b^{22},c^{22})
\end{split}
\end{align}

\noindent Note that there exist additional relations between $Q_n$ and the original distribution $P(A,B,C)$, some of which involve polynomials of the probabilities $P(A,B,C)$ with degree greater than $n$. For instance, 
\begin{equation}\label{eq:higherrelations}
Q_2(\{A^{12}{=}a, B^{12}{=}b, C^{12}{=}c\})=P_A(a)P_B(b)P_C(c).
\end{equation} 
\noindent In this paper we will not exploit such higher degree relations, though they are quite useful in practical implementations.\footnote{For instance, the result in Example~2 of Ref.~\cite{wolfe2016inflation} excluding the so-called \enquote{W-distribution} can be recovered using second order inflation if the higher degree relation of Eq.~\eqref{eq:higherrelations} is imposed. Without that relation, however, the W-distribution is excluded only at \emph{third} order inflation.}

Given an arbitrary distribution $P(A,B,C)$, the \term{inflation technique} consists in demanding the degree-$n$ lifting of $P(A,B,C)$ be the degree-$n$ diagonal marginal of a distribution $Q_n$ over the inflated variables satisfying (\ref{symmetry}). When condition~(\ref{probas}) is met, we call the associated distribution $Q_n$ an \term{\boldmath$n^{th}$-order inflation} of $P$. Clearly, if $P(A,B,C)$ does not admit an $n^{th}$-order inflation for some $n$, then it cannot be realized in the triangle scenario. Deciding if the degree-$n$ lifting of $P(A,B,C)$ is a member of the set of degree-$n$ diagonal marginals can be cast as a linear program~\cite{alevras2001linear}. 

If the linear program is infeasible, i.e., if no $n^{th}$-order inflation exists for $P(A,B,C)$, then the program will find a witness to detect its incompatibility. Such a witness will be of the form 
\be
\overline{F}\cdot P^{\otimes n}<\min_{Q_n}\overline{F}\cdot Q_n^n,
\ee
\noindent where $\overline{F}$ is a real vector and the minimum on the right hand side is taken over all distributions $Q_n$ satisfying Eq.~(\ref{symmetry}). Call $F$ the $n^{th}$-degree polynomial such that $F(Q)=\overline{F}\cdot Q^{\otimes n}$ for all $Q$. For some distributions $P$, the inflation technique will thus output a polynomial witness of incompatiblity $F$, hence solving the corresponding (Approximate) Causal Compatibility problem.

Note that, for $n\geq n'$, any distribution $P$ admitting an $n^{th}$-order inflation also admits an $(n')^{th}$-order inflation. This suggests that we might be able to detect the incompatibility of a distribution $P$ via the inflation technique just by taking the order $n$ high enough.

Since any polynomial of a probability distribution can be lifted to a linear function acting on $g$-liftings, we can also use the inflation technique to attack Approximate Causal Optimization, as long as the function $F$ to minimize happens to be a polynomial. Suppose that this is the case and that $F$ has degree $g$. We wish to minimize $F(P)$ over all distributions compatible with the triangle scenario. Our first step would be to express $F$ as a vector $\overline{F}$, such that $F(P)=\overline{F}\cdot P^{\otimes g}$, for all distributions $P$. Our second step consists in solving the linear program
\begin{align}\begin{split}
&f_n\equiv\min_{Q_n} \overline{F}\cdot Q_n^g,\\
\mbox{where}&\quad Q_n^g \mbox{ is defined by Eq.~(\ref{diagtriangle})},\\
\mbox{and such that}&\quad  Q_n \mbox{ is a distribution}\\ &\quad \mbox{satisfying condition~(\ref{symmetry})}.
\label{optimInflation}
\end{split}\end{align}

Since the $g$-lifting of any distribution $P$ compatible with the triangle scenario can be viewed as the diagonal marginal of degree $g$ of a distribution $Q_n$ satisfying (\ref{symmetry}), we thus have that $f_n\leq f_\star=\min_P F(P)$, for all $n$, just as in the definition of Approximate Causal Optimization. Moreover, $f_n\geq f_{n'}$, for $n\geq n'$, i.e., as we increase the order $n$ of the inflation, we should expect to obtain increasingly tighter lower bounds on $f_\star$. If, by whatever means, we were to obtain an upper bound $f_+$, then we would have solved Approximate Causal Optimization for all $\epsilon\geq f_+-f_n$.

In the triangle scenario, the inflation technique can therefore be used to tackle both Approximate Causal Compatibility and Approximate Causal Optimization.

\begin{figure}[h]
  \centering
  \includegraphics[scale=0.8]{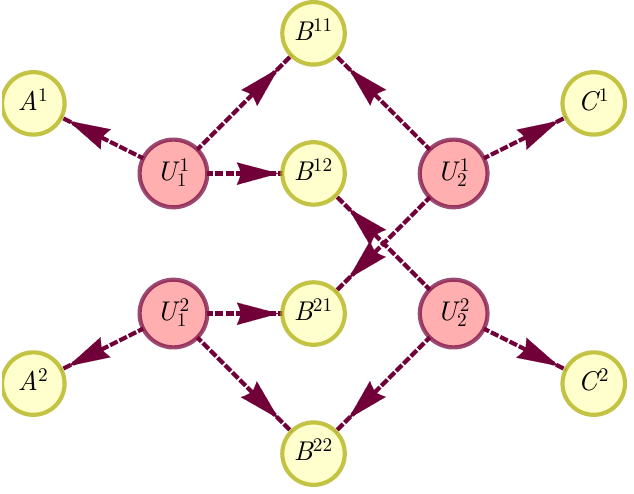}
  \caption{\textbf{The second-order inflation graph of the three-on-line scenario.}}\label{fig:ThreeLineInflated}
\end{figure}

For further elucidation, consider another correlation scenario. In the \term{three-on-line scenario}, Fig.~\ref{bilocality}, we again have three random variables $A,B,C$ which are defined, respectively, via the non-deterministic functions $A(U_1), B(U_1,U_2), C(U_2)$. As always, the exogenous latent variables $\{U_1,U_2\}$ are independently distributed. The ${n=2}$ inflation graph for the three-on-line scenario is depicted in Fig.~\ref{fig:ThreeLineInflated}.

In this scenario, an $n^{th}$-order inflation corresponds to a distribution $Q_n$ over the variables $\left\{\{A^{i}\}, \{B^{jk}\}, \{C^l\}\right\}$, where $i,j,k,l$ range from $1$ to $n$. $Q_n$ must satisfy the linear constraints:
\begin{align}\begin{split}
&Q_n(\{A^{i}{=}a^{i}, B^{jk}{=}b^{jk}, C^{l}{=}c^{l}\})\\
&=Q_n(\{A^{i}{=}a^{\pi(i)},B^{jk}{=}b^{\pi(j)\pi'(k)}, C^{l}{=}c^{\pi'(l)}\}),
\end{split}\label{symmetry2}
\end{align}
\noindent for all permutations of $n$ elements $\pi,\pi'$. Expanded out for $n=2$, Eq.~\eqref{symmetry2} becomes \begin{subequations}
\begin{align}
&\nonumber Q_2(a^1,a^2,b^{11},b^{12},b^{21},b^{22},c^1,c^2)\\
&= Q_2(a^2,a^1,b^{21},b^{22},b^{11},b^{12},c^1,c^2)\\
&= Q_2(a^1,a^2,b^{12},b^{11},b^{22},b^{21},c^2,c^1)\label{eq:bilocalitysymmetry2}
\end{align}
\end{subequations}
Additionally, relating the degree-$g$ liftings of $P$ to the diagonal marginal in this scenario requires
\begin{align}
\hspace{-2ex}Q^g_n\big(\bigwedge\limits_{i=1}^g \{A^{i}{=}a^i,B^{ii}{=}b^{i},C^{i}{=}c^i\}\big)=\prod_{i=1}^g P(a^i,b^i,c^i),
\label{probas2}
\end{align}
for any choice of integer $g$ such that ${g\leq n}$. Expanded out for ${g{=}n{=}2}$, Eq.~\eqref{probas2} identifies the diagonal marginal for this scenario ${Q_{n{=}2}^{g{=}2}={\sum_{B^{12},B^{21}}}
Q_{n{=}2}}$ as 
\begin{align}\begin{split}\label{eq:bilocalityidentify}
Q_{n{=}2}^{g{=}2}(a^1,&a^2,b^{11},b^{22},c^1,c^2)\\
&=P(a^1,b^{11},c^1)P(a^2,b^{22},c^{2})\\
\end{split}
\end{align}

The above ideas are easy to generalize to arbitrary correlation scenarios (remember, though, that correlation scenarios are just a special class of causal structures). 

\subsection{Inflation of an Arbitrary Correlation Scenario}
To set up the  $n^{th}$-order inflation of an arbitrary correlation scenario, first imagine $n$ independent copies of all the latent variables, and then consider all the observable variables which are children of these, following the prescription of the original correlation scenario. Each observable variable in the inflation graph has as many superindices as latent variables it depends on. Then, one must impose symmetry restrictions on the total probability distribution $Q_n$, demanding that it be invariant under any relabeling-permutations applied to the index of any one latent variable, i.e.,
\begin{align}\label{symmetry^gen}
&Q_n(\{A_1^{\bar{i}_1}{=}a_1^{\bar{i}_1}{,}...,A_m^{\bar{i}_m}{=}a_m^{\bar{i}_m}:\bar{i}_1{,}...,\bar{i}_m\})=\\\nonumber
&Q_n(\{A_1^{\bar{i}_1}{=}a_1^{\bar{\pi}^{L_1}(\bar{i}_1)}{,}...,A_m^{\bar{i}_m}{=}a_m^{\bar{\pi}^{L_m}(\bar{i}_m)}:\bar{i}_1{,}...,\bar{i}_m\}),
\end{align}
\noindent for all vectors $\bar{\pi}=(\pi^1{,}...,\pi^L)$ of $L$ independent permutations (one for each latent variable or index type). Here $\bar{i}_x$ denotes the tuple of superindices on which variable $A_x$ depends. 
It should go almost without saying that we demand non-negativity and normalization of the inflation probabilities
\begin{align}\begin{split}\label{positivityofinflation}
&Q_n(\{A_1^{\bar{i}_1}{=}a_1^{\bar{i}_1}{,}...,A_m^{\bar{i}_m}{=}a_m^{\bar{i}_m}:\bar{i}_1{,}...,\bar{i}_m\})\geq 0,\\
&\sum_{\vec{a}}Q_n(\vec{a})=1.
\end{split}\end{align}
The central object we consider, then, is the set of all diagonal marginals consistent with such an $n^{th}$-order inflation. We denote such a generic diagonal marginal by
\begin{align}
&Q^g_n\coloneqq\; Q_n \big(\bigwedge\limits_{i=1}^g\{A_1^{i...i}{=}a_1^i{,}...,A_m^{i...i}{=}a_m^i\}\big).
\label{diag^gen}\end{align}
The compatibility conditions 
\begin{align}
&\hspace{-1ex}Q_n^g\big(\bigwedge\limits_{i=1}^g\{A_1^{i...i}{=}a_1^i{,}...,A_m^{i...i}{=}a_m^i\}\big){=}\prod_{i=1}^g P(a_1^i{,}...,a_m^i)
\label{probas^gen}
\end{align}
require the degree-$g$ lifting of $P$ to be consistent with such a degree-$g$ diagonal marginal.

Notice that 
any distribution $Q_n$ subject to the constraints (\ref{symmetry^gen}-\ref{probas^gen}) must be such that the marginals associated with relabellings of the indices of the diagonal variables obey the same compatibility conditions as the canonical diagonal marginals do, i.e.,
\begin{align}\begin{split}
Q_n\big(&\bigwedge\limits_{i=1}^g \{A_1^{\bar{\pi}^{L_1}(i...i)}{=}a_1^i{,}...,A_m^{\bar{\pi}^{L_m}(i...i)}{=}a_m^i\}\big)\\
&=\prod_{i=1}^g P(a_1^i{,}...,a_m^i),
\end{split}\label{probas^wolfe}
\end{align}
\noindent for all $\bar{\pi}$. 

Actually, the original description of the inflation technique in Ref.~\cite{wolfe2016inflation} imposes the constraints (\ref{probas^wolfe}) rather than (\ref{symmetry^gen}-\ref{probas^gen}) over the distribution $Q_n$, as 
demanding the existence of a distribution $Q_n$ satisfying condition (\ref{probas^wolfe}) can be shown to enforce over $P(a_1{,}...,a_m)$ exactly the same constraints as demanding the existence of a distribution satisfying (\ref{symmetry^gen}-\ref{probas^gen}). Indeed, as noted in Ref.~\citep[App.~C]{wolfe2016inflation}, any distribution $Q_n$ satisfying (\ref{probas^wolfe}) can be twirled or symmetrized (see Appendix) to a distribution $\tilde{Q}_n$ satisfying Eqs.~(\ref{symmetry^gen}-\ref{probas^gen}). For convenience, from now on we will just refer to the formulation of the inflation technique involving the symmetries (\ref{symmetry^gen}). This formulation has the added advantage that the symmetry constraints can be exploited to reduce the time and memory complexity of the corresponding linear program, see for instance Ref.~\cite{Gent2006329}.

It isn't hard to see how this general notion of inflation can also be used to tackle Approximate Causal Compatibility and Approximate Causal Optimization in general correlation scenarios $\G$:

\begin{prob}{\term{Inflation for Causal Compatibility}}\label{Infl^Comp}\\
\noindent\textsc{Input:} A positive integer $n$, a causal structure $\G$ and a particular probability distribution over the observed variables $P$.\\
\noindent\textsc{Primal Linear Program:} 
\begin{align}\begin{split}
&\min_{Q_n} 0,\\
\mbox{where}&\quad P\text{ relates to }Q_n \mbox{ by Eqs.~(\ref{diag^gen},\ref{probas^gen})},\\
\mbox{and such that}&\quad Q_n \mbox{ satisfies conditions~(\ref{symmetry^gen},\ref{positivityofinflation})}.
\end{split}\end{align}
\noindent\textsc{Dual Linear Program:} 
\begin{align}\begin{split}
&\min_{\overline{F}} \overline{F}\cdot P^{\otimes n},\\
\mbox{such that}&\quad 0\leq \overline{F}\cdot Q_n^{g{=}n} \leq 1,\\
\mbox{where}&\quad Q_n^{g} \mbox{ is defined by Eq.~(\ref{diag^gen})},\\
\mbox{and such that}&\quad Q_n \mbox{ satisfies conditions~(\ref{symmetry^gen},\ref{positivityofinflation})}.
\end{split}\end{align}
\noindent\textsc{Summary:} 
If the degree-$n$ lifting of $P$ is not in the set of degree-$n$ diagonal marginals consistent with an $n^{th}$-order inflation of $\G$, then the returned dual variable $\overline{F}$ will witness the incompatibility of $P$ per $\overline{F}\cdot P^{\otimes n}< 0$ while $\overline{F}\cdot P'^{\otimes n}\geq 0$ for all distributions $P'$ compatible with $\G$.
\end{prob}

Similarly,

\begin{prob}{\term{Inflation for Causal Optimization}}\label{Infl^Opt}\\
\noindent\textsc{Input:} A positive integer $n$, a causal structure $\G$ and a degree-$g$ polynomial function $F$ of the probabilities of the observed events.\\
\noindent\textsc{Linear Program:} 
\begin{align}\begin{split}
f_n\equiv\quad&\min_{Q_n} \overline{F}\cdot Q_n^g,\\
\mbox{where}&\quad Q_n^{g} \mbox{ is defined by Eq.~(\ref{diag^gen})},\\
\mbox{and such that}&\quad Q_n \mbox{ satisfies conditions~(\ref{symmetry^gen},\ref{positivityofinflation})}.
\end{split}\end{align}
\noindent\textsc{Summary:} 
The programs returns a degree-$g$ diagonal marginals consistent with an $n^{th}$-order inflation of $\G$ which minimizes the input function $F$. Since such diagonal marginals contain all degree-$g$ liftings of distributions compatible with $\G$, it follows that $f_n$ is a lower bound on the minimum value of $F$ over all distributions compatible with $\G$.
\end{prob}

%

In certain practical cases, we may not know the full probability distribution of the observable variables, but only the probabilities of a restricted set $E$ of \emph{observable events}. As we will see in Section~\ref{extension}, this often happens when we map the causal compatibility problem from a general causal structure to a correlation scenario. To apply the inflation technique to those cases, rather than fixing the value of all probability products, like in Eq.~(\ref{probas^gen}), we will impose the constraints
\begin{align}\begin{split}
&\smashoperator{\sum_{\vec{a}^1\in e^1{,}...,\vec{a}^n\in e^n}} Q_n(\{A_1^{i...i}{=}a_1^i{,}...,A_m^{i...i}{=}a_m^i\}_i)\\
&=\prod_{i=1}^n P(e^i),
\label{probas^gen2}
\end{split}\end{align}
\noindent for all $e^1{,}...,e^n\in E$. Any distribution $Q_n$ satisfying both (\ref{symmetry^gen}) and (\ref{probas^gen2}) will be dubbed an \emph{$n^{th}$ order inflation of the distribution of observable events}.

For example, consider again the three-on-line scenario (Fig. \ref{bilocality}), and assume that our experimental setup just allows us to detect events of the form $e(a)\equiv\{(A,B,C):A=B=C=a\}$. Then our set of observable events is $E=\cup_a \{e(a)\}$ and the input of the causal inference problem is the distribution $\{P(e), e\in E\}$. An $n^{th}$ order inflation $Q_n$ of $P(e)$ would satisfy Eq.~(\ref{symmetry2}) and the linear conditions
\begin{align}\begin{split}
&Q_n(\{A^{i}{=}a^i,B^{ii}{=}a^i,C^{i}{=}a^i\}_i)=\prod_{i=1}^n P(e(a^i))\\
&=\prod_{i=1}^n P(a^i,a^i,a^i).
\label{probas3}
\end{split}\end{align}

\section{Convergence of Inflation}\label{convergencenoproof}
The main result of this article is that the inflation technique can be used to solve Approximate Causal Compatibility and Approximate Causal Optimization for arbitrarily small values of $\epsilon$, just by taking the order $n$ of the inflation high enough. Depending on which of the two problems we wish to solve and which causal structures are involved, we will have either finite-order convergence or asymptotic convergence.

\subsection{On finite-order convergence}

Even at low orders, the Inflation Technique has been shown to provide very good outer approximations to the set of distributions compatible with the triangle scenario \cite{wolfe2016inflation}. Furthermore, for certain correlation scenarios, a second-order inflation can be shown to fully characterize the set of compatible distributions.

Consider, for instance, the three-on-line scenario (Fig.~\ref{bilocality}), whose second-order inflation was depicted in Fig.~\ref{fig:ThreeLineInflated}. Note that condition~(\ref{eq:bilocalityidentify}) implies that
\begin{align}
Q_{n{=}2}^{g{=}2}(A^1{=}a^1,C^2{=}c^2)=P(A{=}a^1)P(C{=}c^{2})
\label{extra1}
\end{align}
and condition~(\ref{eq:bilocalitysymmetry2}) implies that
\begin{align}\begin{split}
&Q_{n{=}2}(A^1{=}a^1,C^1{=}c^1,C^2{=}c^2) 
\\&= Q_{n{=}2}(A^1{=}a^1,C^1{=}c^2,C^2{=}c^1).
\end{split}
\label{extra2}
\end{align}
From the last condition, it follows that ${Q_{n=2}(A^1{=}a, C^1{=}c)}={Q_{n=2}(A^1{=}a, C^2{=}c)}$. Invoking condition~(\ref{extra1}), we thus have that
\begin{align}
&Q_{n{=}2}(A^1{=}a^1,C^1{=}c^1)
=P(A{=}a^1)P(C{=}c^1).
\end{align}
This is sufficient to ensure that $Q_{n{=}2}^{g{=}1}$ is realizable in the three-on-line scenario, since then $P(A,B,C)=P(A,C)P(B|A,C)=P(A)P(C)P(B|A,C)$. This last expression represents a realization of $P(A,B,C)$ in the three-on-line scenario, where the hidden variables $U_1,U_2$ are, respectively, $A$ and $C$.

\begin{figure}[h]
  \centering
  \includegraphics[scale=0.8]{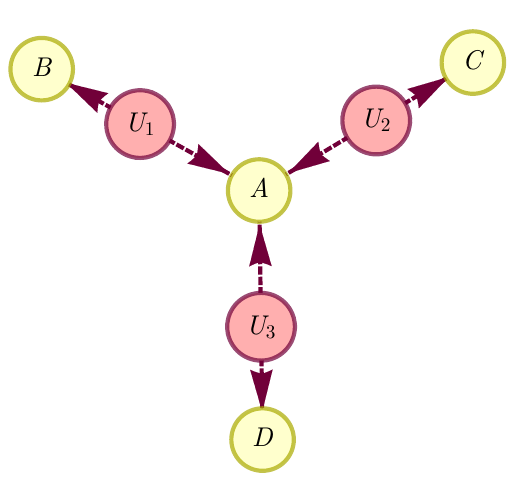}
  \caption{\textbf{A star-shaped correlation scenario.}}
  \label{fig:Star}
\end{figure}

This example can be generalized to prove convergence at order ${n{=}2}$ of any star-shaped correlation scenario. Star-shaped scenarios with $N$ observable variables have the defining property that, in some subset of ${N{-}1}$ observable variables, every pair of variables share no latent parents~\cite{Tavakoli2017BellToNetwork,Chaves2017starnetworks}, see Figs.~\ref{bilocality}~and~\ref{fig:Star}. (This definition assumes that every set of variables in a correlation scenario all of which have the \emph{same} set of latent parents are implicitly merged into a single vector-value variable.) Given an arbitrary star-shaped correlation scenario with $N$ random variables, call $B_1{,}...,B_{N-1}$ any set of ${N{-}1}$ random variables without a common ancestor; and $A$, the remaining variable. Using the same trick as before, one can prove that, for any ${i{\neq}j}$, $P(B_i,B_j)=P(B_i)P(B_j)$. Similarly, one can group variables $B_i,B_j$ and argue that, for any ${l{\neq}i},j$, $P(B_i,B_j,B_l)=P(B_i,B_j)P(B_l)=P(B_i)P(B_j)P(B_l)$. Iterating this argument, we show that ${P(B_1{,}...,B_{N-1})}$ factors into ${N{-}1}$ products. Analogously, it is proven that $P(A,B_{i_1}{,}...,B_{i_m})=P(A)P(B_{i_1}{,}...,B_{i_m})$, for any set of indices ${i_1{,}...,i_m}$ such that ${B_{i_1}{,}...,B_{i_m}}$ do not share parents with $A$. This is enough to prove compatibility.

\begin{figure}[t]
  \centering
  \includegraphics[scale=0.8]{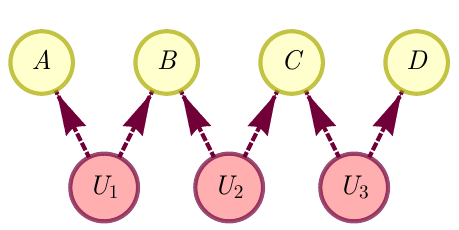}
  \caption{\textbf{The {four-on-line} correlation scenario.}}
  \label{fig:FourLine}
\end{figure}

In these examples, using the inflation technique is an overkill, as compatibility can be determined solely by checking the satisfaction of all independence relations. There are many correlation scenarios, however, where distribution compatibility is also determined by \emph{inequality} constraints. Examples of such \enquote{interesting}\footnote{Here we use \enquote{interesting} in precisely the meaning of Refs.~\cite{HLP,Pienaar2017}.} correlations scenarios include the triangle scenario, as well as the four-on-line scenario depicted in Fig.~\ref{fig:FourLine}. And actually, in the former scenario, the problem of compatibility of distributions is \emph{not} completely solved by second-order inflation.

Indeed, \emph{all} binary variable distributions of the form
\begin{align}
P_v(A{=}a,B{=}b,C{=}c)=\begin{cases}
v &\text{if } abc=111\\
\frac{1-v}{7} &\text{otherwise.}\end{cases}
\end{align}
pass the second-order inflation test for triangle scenario causal compatibility. On the other hand, the Finner inequality applied to the triangle scenario in Ref.~\citep[Thm.~1]{Renou2019networks} certifies the incompatibility of all $P_v$ for which $v>57/64$.

To be clear, however, there exists situations where, conversely, a second order inflation outperforms the Finner inequality. For probability distributions of the form
\begin{align}
P_{q,r}(A{=}a,B{=}b,C{=}c)=\begin{cases}
q &\text{if } abc=000\\
r &\text{if } abc=111\\
\frac{1-q-r}{6} &\text{otherwise.}\end{cases}
\end{align}
compatibility with second order inflation for the triangle scenario requires
\begin{align}
q\leq \frac{13r+8-3\sqrt{16r(r{+}2)+1}}{5}
\end{align}
This bound is strictly tighter than bounds implied by the Finner inequality~\cite{Renou2019networks} or the semidefinite causal compatibility constraints involving covariances of Refs.~\cite{Kela2017Covariance,Aberg2020semidefinite} throughout the parameter region $0.0283\lesssim r\leq q$.

%


For arbitrary correlation scenarios $\G$ with observed variables of specified cardinality, we inquire whether \emph{some} finite-order inflation is always sufficient to characterize the set of compatible distributions. Are there causal structures for which inflation converges only asymptotically? Could the triangle scenario be such an example? 

\begin{openq}
For any correlation scenario $\G$, does there exist $n$ such that $n^{th}$-order inflation solves exact Causal Compatibility? 
\end{openq}

Interestingly, we can prove that, when used to solve Approximate Causal \emph{Optimization}, the inflation technique does not converge, in general, in a finite number of steps. Indeed, consider the trivial correlation scenario consisting of a single observable variable $A$ and its single latent parent $U$. We wish to use the inflation technique to minimize the function $-P(A{=}0)P(A{=}1)$. Clearly the solution of this problem is $-\frac{1}{4}$. An $n^{th}$-order function inflation assessment (starting at $n\geq 2$), however, would effectively reduce this problem to the LP
\begin{align*}\begin{split}
\hspace{-5ex}\smashoperator{\min_{Q_n}} \quad &-Q_n^{g{=}2}(0,1)\equiv -\smashoperator{\sum_{a^3{,}...,a^n}}Q_n(0,1,a^3{,}...,a^n)\quad\text{s.t.}\\
&Q_n(a^1{,}...,a^n)\geq 0,\quad
\smashoperator[r]{\sum_{a^1{,}...,a^n}}Q_n(a^1{,}...,a^n)=1,\\
&Q_n(a^1{,}...,a^n)=Q_n(a^{\pi(1)}{,}...,a^{\pi(n)}),\;\forall\pi\in S^n.
\end{split}\end{align*}
\noindent For $n=2n'$, consider the symmetric probability distribution $Q_n$ given by randomly choosing without replacement $n$ bits $a^1{,}...,a^n$ from a pool of $n'$ $0$'s and $n'$ $1$'s. Then it can be verified that 
\be
-\smashoperator{\sum_{a^3{,}...,a^n}}Q_n(0,1,a^3{,}...,a^n)=-\frac{1}{2}\frac{n'}{2n'-1}<-\frac{1}{4},
\ee
\noindent overshooting the magnitude of the true minimum for all $n'$. Nonetheless, note that the above quantity converges to the correct result of $-\frac{1}{4}$ asymptotically as $O(1/n)$.

\subsection{Asymptotic convergence}
In this section we will prove that, for any correlation scenario $\G$, the inflation technique characterizes the set of compatible correlations asymptotically. More precisely, we will show that any distribution $P$ admitting an $n^{th}$ order inflation is $O(1/\sqrt{n})$-close in Euclidean norm to a compatible distribution $\tilde{P}$. Similarly, we will show that $f_n$, as defined in Eq.~(\ref{optimInflation}), satisfies $f_\star-f_n\leq O(1/n)$. In order to solve Approximate Causal Compatibility and Approximate Causal Optimization for a given value of $\epsilon$, we just have to use the Inflation Technique up to orders $O(1/\epsilon^2)$, $O(1/\epsilon)$, respectively. Since the set of compatible distributions is closed \cite{rosset2016finite}, this implies that, for any incompatible distribution $P$, there exists $n$ such that $P$ does not admit an $n^{th}$ order inflation.

Before we proceed with the proof, a note on the scope of our results is in order. The inflation technique is fairly expensive in terms of time and memory resources. At order $n$, it involves optimizing over probability distributions of $\sum_{i=1}^m n^{|L_i|}$ variables (we remind the reader that $L_i\subset\{1,...,L\}$ denotes the set of indices $j$ such that the hidden variable $U_j$ influences $A_i$). If each of these variables can take $d$ possible values, then the number of free variables in the corresponding linear program is $N\equiv d^{\sum_{i=1}^m n^{|L_i|}}$. That is, the memory resources required by the inflation technique are superexponential on $n$. Add to this the fact that the best LP solvers in the market have a running time of $O(N^3)$ \cite{LPcompl}, and you will come to the conclusion that a brute-force implementation of the inflation technique in the triangle scenario is already unrealistic for $n=4$, even in the simplest case of $d=2$. What is the relevance, then, of proving asymptotic convergence? 

For us, it is a matter of principle. Even at low orders, the inflation technique has proven itself very useful at identifying non-trivial constraints on observable probability distributions. It is therefore natural to ask whether the inflation technique just provides a partial characterization of compatibility, or, on the contrary, it reflects an alternative way of comprehending the latter. Our work settles this question completely: by proving that any unfeasible distribution must violate one of the inflation conditions, we refute the first hypothesis and validate the second one.

The key to deriving the asymptotic convergence of the Inflation Technique is the following theorem, proven in the Appendix.

\begin{theo}\label{ext^definetti}
Let $\G$ be a correlation scenario with $L$ latent variables, and let $Q_n^g$ be the degree-$g$ diagonal marginal of a distribution $Q_n$ satisfying the symmetry conditions~(\ref{symmetry^gen}). Then, there exist normalized probability distributions $P_\mu$ compatible with $\G$ and probabilities $p_\mu\geq 0$, $\sum_\mu p_\mu=1$ such that
\be
{D}{\left(Q_n^g,\sum_\mu p_\mu P_\mu^{\otimes g}\right)}\leq {O}{\left(\frac{Lg^2}{n}\right)},
\ee
\noindent where $D(q,r)=\sum_x |q(x)-r(x)|$ denotes the \emph{total variation distance} between the probability distributions $q(X),r(X)$.
\end{theo}

This theorem can be regarded as an extension of the finite de Finetti theorem \cite{deFinetti}, that states that the marginal $P(a^1{,}...,a^g)$ of a symmetric distribution $P(a^1{,}...,a^n)$ is $O(g^2/n)$-close in total variation distance to a convex combination of degree-$g$ liftings.

The solvability of Approximate Causal Optimization through the Inflation Technique follows straightforwardly from Theorem \ref{ext^definetti}. Let $F$ be a polynomial of degree $g$, with $f_\star=\max_P F(P)$, and let $Q_n$ be the symmetric distribution achieving the value $f_n$ in Eq.~(\ref{optimInflation}). Then, by the previous theorem, we have that 
\begin{align}\begin{split}
f_n=\overline{F}\cdot Q^g_n&= \sum_{\mu}p_\mu \overline{F}\cdot P^{\otimes g}_\mu-{O}{\left(\frac{Lg^2}{n}\right)}\\
&=\sum_{\mu}p_\mu F(P_\mu)-{O}{\left(\frac{Lg^2}{n}\right)}\\
&\geq f_\star-{O}{\left(\frac{Lg^2}{n}\right)}.
\end{split}
\shortintertext{It follows that}
 f_n\leq f_\star&\leq f_n +{O}{\left(\frac{Lg^2}{n}\right)}.
\label{demoFin}
\end{align}

Proving the analog result for Approximate Causal Compatibility is only slightly more complicated. Let $P$ be a probability distribution over the observed variables, and suppose that $P$ admits an $n^{th}$-order inflation $Q_n$. Define the second-degree polynomial $N(R)=\sum_{\bar{a}} (R(\bar{a})-P(\bar{a}))^2$, and let $\overline{N}$ be a linear functional such that $\overline{N}\cdot q^{\otimes 2}=N(q)$ for all distributions $q$. Note that, due to conditions~(\ref{probas^gen}), $\overline{N}\cdot Q^2_n=N(P)=0$. Thus the minimum value $f_n$ of $\overline{N}\cdot Q^2_n$ over all diagonal marginals of degree $2$ of a distribution $Q_n$ satisfying the symmetry conditions~(\ref{symmetry^gen}) is such that $f_n\leq 0$. Invoking Eq.~(\ref{demoFin}) for $g=2$, we have that
\begin{align}
f_n\leq f_\star\leq f_n+{O}{\left(\frac{L}{n}\right)},
\label{deriva}
\end{align}
\noindent where $f_\star$ is the minimum value of $N(Q)$ over all compatible distributions $Q$. This implies that there exists a compatible distribution $\tilde{P}$ such that 
\be
\sqrt{N(\tilde{P})}=\|P-\tilde{P}\|_2\leq {O}{\left(\sqrt{\frac{L}{n}}\right)}.
\label{cosicaGuapa}
\ee

This proof of convergence easily extends to scenarios where we only know the probabilities of set $E$ of observable events. Indeed, choosing the polynomial $N$ such that $N(R)=\smashoperator{\sum_{e\in E}} \left(P(e)-R(e)\right)^2$, and following the same derivation as in Eq.~(\ref{deriva}), we conclude that a distribution of observable events admitting an $n^{th}$ order inflation is ${O}{\left(\sqrt{\frac{L}{n}}\right)}$-close in Euclidean norm to a compatible distribution.

\section{Unpacking Causal Structures}
\label{extension}

So far we have just been referring to correlation scenarios, i.e., those causal structures where all observed variables only depend on a number of independent latent variables. However, in a general causal model, the value of a given variable can depend, not only on latent variables, but also on the values of other observed variables. In the following, we define procedures call \term{exogenization} and \term{unpacking} which cumulatively map the problem of causal compatibility with an arbitrary causal structure to problems of causal compatibility with the structure's implicit constituent correlation scenarios. Consequently, these procedures enable application of the inflation technique to general causal structure via preprocessing into correlation scenarios.

Suppose $\G$ is a DAG with latent variables. If $U$ is an endogenous (non-root) latent variable in $\G$, one can \term{exogenize} $U$ by first adding all possible directed edges originating from a parent of $U$ and terminating at a child of $U$, and then deleting from $\G$ all directed edges which terminated at $U$. The resulting graph admits precisely the same set of feasible observed distributions as $\G$, per Ref.~\cite[Sec.~3.2]{evans}. Hereafter, therefore, we restrict our attention to causal compatibility problems involving causal structures where all latent variables are exogenous.

%
%
%
%
%
%

\begin{figure}[t]
  \centering
  \includegraphics[scale=0.8]{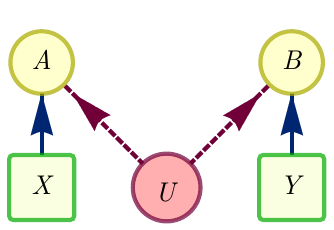}
  \caption{\vtop{\hbox{\textbf{The Bell scenario.}}(Not a correlation scenario.)}}
  \label{fig:Bell}
\end{figure}

In addition, we will always consider distributions as implicitly conditional on the values of any exogenous observable variables. Of course, this mapping from raw probability distributions to conditional probability distributions only makes sense if the distribution of exogenous observable variables factorizes, i.e., if all exogenous observable variables are independent from each other. As an example, the sorts of distributions we consider for the Bell scenario depicted in Fig.~\ref{fig:Bell} are of the form $P(A,B|X,Y)$, as opposed to $P(A,B,X,Y)$.

To go from general causal structures to correlation scenarios, we introduce counterfactual variable sets, in which we consider all the different ways a variable can respond to its observable parents as distinct variables. We call the procedure for eliminating all dependencies between observed variables \term{unpacking}. 
Unpacking is related to --- but distinct from --- the \emph{single world intervention graphs} introduced in Ref.~\cite{Richardson2013SingleWI} and the \emph{e-separation} method developed in Ref.~\cite{evans2012graphicalmethods}. As quantum physicists, we understand unpacking as a manifestation of \emph{counterfactual definiteness}, which is a natural assumption mysteriously inconsistent with quantum theory~\cite{gill2014,Leifer2005PPS,Liang2011}. Since counterfactual definiteness \emph{does} hold in the ``classical'' causal models considered in this paper, we promote maximally exploiting this assumption as a first step towards resolving any causal compatibility problem.

\begin{figure}
  \centering
  \includegraphics[scale=0.8]{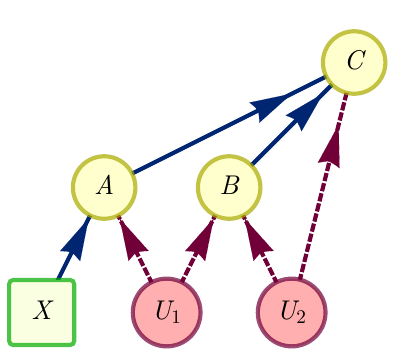}
  \caption{\vtop{\hbox{\textbf{The example structure \boldmath$\G^1$.}}(Not a correlation scenario.)}}
  \label{fig:ESepExample}
 \centering
 \includegraphics[scale=0.8]{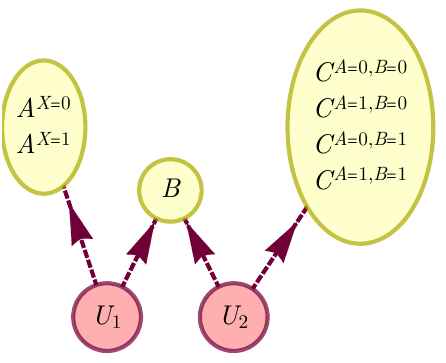}
 \caption{\textbf{The unpacking of the example structure \boldmath$\G^1$, which we denote by $\G^2$.}}
   \label{fig:ESepUnpacked}
\end{figure}

By way of example, consider the structure $\G^1$ depicted in Fig.~\ref{fig:ESepExample}. 
The correlation scenario which results from unpacking $\G^1$ --- assuming that all observable variables are discretely valued in the range $[0,1]$ --- is shown in Fig.~\ref{fig:ESepUnpacked}. The unpacked scenario can be though of as having either seven binary-valued variables $\{A^{X{=}0}$, $A^{X{=}1}$, $B$, $C^{A{=}0,B{=}0}$, $C^{A{=}0,B{=}1}$, $C^{A{=}1,B{=}0}$, $C^{A{=}1,B{=}1}\}$ or simply three variables, two of which are vector valued. We use the latter interpretation for the visualization of the unpacked scenario, but the former interpretation is convenient to explicitly relate the packed distributions to the unpacked distributions. A distribution over the observable variables in $\G^1$ (conditioned on the exogenous observable variable $X$) is compatible with $\G^1$ iff there exists another distribution compatible with $\G^2$ (over $\G^2$'s observable variables) such that the first distribution is recovered via suitable \emph{varying} marginals of the latter. Explicitly,
\begin{align}\begin{split}
&P_{\text{original}}(A{=}a,B{=}b,C{=}c|X{=}x)\\ 
&=P_{\text{unpacked}}(A^{X{=}x}{=}a,
	B{=}b,
	C^{A{=}a,B{=}b}{=}c).
	\label{eq:esepexplicit}
\end{split}\end{align}
Note that for each of the eight distinct choices of $\{a,b,x\}\in\{0,1\}^3$, the marginal of $P_{\text{unpacked}}$ referenced in Eq.~\eqref{eq:esepexplicit} is distinct. The \emph{subset of variables} specifying the relevant marginal of $P_{\text{unpacked}}$ does \emph{not} vary depending on the value of $c$, however.

We now describe how to unpack an arbitrary causal structure. Let $A$ be an observed variable. We denote the observable parents of $A$ as $\textsf{paOBS}[A]$. Suppose the (set) $\textsf{paOBS}[A]$ can take $d$ different values, e.g.: $\textsf{paOBS}[A]\in\{1{,}...,d\}$. The different values of $\textsf{paOBS}[A]$ are generally vector-valued; we may nevertheless denote such value-tuples by a single scalar index, for compactness of notation.  To unpack the vertex $A$, we break all edges between $\textsf{paOBS}[A]$ and $A$, unpacking $A$ into the counterfactual variable set $\{A^{\textsf{paOBS}[A]{=}1}{,}...,A^{\textsf{paOBS}[A]{=}d}\}$. Unpacking all the endogenous observed variables, and regarding the resulting counterfactuals as observed variables themselves, we arrive at the associated correlation scenario.

The probabilities of the observed variables in $\G$ can be obtained from the probabilities of a set of measurable events in the associated correlation scenario $\G'$. To be clear, let $\bar{A}$ ($\bar{X}$) denote all the observable endogenous (exogenous) variables in $\G$. Then,
\begin{align}\begin{split}\label{eq:unpacking}
&P_{\text{original}}(\bar{A}{=}\bar{a}|\bar{X}{=}\bar{x})
\\&=P_{\text{unpacked}}(\bigwedge_i A_i^{\textsf{paOBS}[A_i]{=}\{\bar{a},\bar{x}\}_{A_i}}{=}a_i)
\end{split}\end{align}
where $\{\bar{a},\bar{x}\}_{A_i}$ denotes selecting those elements out of the set $\bar{a}\cup\bar{x}$ which corresponding to the values of $\textsf{paOBS}[A_i]$. 

The original Approximate Causal Compatibility (Approximate Causal Optimization) problem in $\G$ is thus mapped to an Approximate Causal Compatibility (Approximate Causal Optimization) problem in the correlation scenario $\G'$, with a non-trivial set of observable events. The inflation technique can then be applied on $\G'$ to solve either problem on the original structure $\G$ up to arbitrary precision $\epsilon$.

\begin{figure}[b]
 \centering
 \includegraphics[scale=0.8]{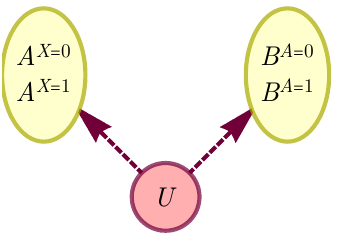}
 \caption{\textbf{The unpacking of the instrumental scenario.}}\label{fig:InstUnpacked}
\end{figure}

Note that unpacking can be valuable even without further inflation. For instance, unpacking the instrumental scenario of Fig.~\ref{instrumental} leads to an associated correlation scenario which is trivial, such as depicted in Fig.~\ref{fig:InstUnpacked}. \emph{Any} distribution over the four variables $\{A^{X{=}0},A^{X{=}1},B^{A{=}0},B^{A{=}1}\}$ is compatible with Fig.~\ref{fig:InstUnpacked}. Nevertheless, demanding that the distributions $P(A,B|X)$ admit such an unpacking leads to nontrivial constraints. We can formulate the admission of an unpacked distribution as a linear program, via Eq.~\eqref{eq:unpacking}. Explicitly formulating this linear program for the instrumental scenario looks like
\begin{align*}\begin{split}\label{eq:instunpacking}
&P_{\text{original}}(A{=}a,B{=}b|X{=}x)=P_{\text{unpacked}}(A^{x}{=}a,B^{a}{=}b),
\\
&P_{\text{unpacked}}(A^{0}{=}a^0,A^{1}{=}a^1,B^{0}{=}b^0,B^{1}{=}b^1)\geq 0,\\
&\smashoperator[r]{\sum_{a^0a^1b^0b^1}} P_{\text{unpacked}}(A^{0}{=}a^0,A^{1}{=}a^1,B^{0}{=}b^0,B^{1}{=}b^1)=1,
\end{split}\end{align*}
and leads to the famous instrumental inequalities~\cite{Pearl95Instrumental}, such as
\begin{align}
P(A{=}0,B{=}0|X{=}0)+P(A{=}0,B{=}1|X{=}1)\leq 1.
\end{align}

This example substantially generalizes. Unpacking alone also completely solves the causal compatibility problem for any single-district causal structure (see \cite{Evans2018NMP} for a definition) containing one (or fewer) latent variables. This includes all Bell scenarios and the entire hierarchy of their relaxations as described in Ref.~\cite{Chaves2017Hieararchy}. 

Furthermore, one can take advantage of known results concerning \term{observationally equivalent} causal structures. We say that $\G$ and $\G'$ are observationally equivalent whenever both structures admit precisely the same set of compatible distributions over their observed variables.
Prop.~5 in Ref.~\cite{evans}
 is a prescription for replacing latent variables with sets of directed edges while preserving observational equivalence. We encourage aggressive application of that prescription in order to convert (unpacked) correlation scenarios into observationally equivalent structures which can be unpacked further. 
For instance, it can be invoked to convert the four-on-line correlation scenario of Fig.~\ref{fig:FourLine} into the observationally equivalent Bell scenario of Fig.~\ref{fig:Bell}, to convert the three-on-line correlation scenario of Fig.~\ref{bilocality} into the observationally equivalent graph $A\rightarrow B\leftarrow C$ with no latent variables, or to replace all latent variables in star scenarios such as Fig.~\ref{fig:Star} with inwards-pointing directed edges. Interestingly, all the challenging causal structures collected in Fig.~14 of Ref.~\cite{evans} unpack to the four-on-line correlation scenario. One can readily demonstrate the non-saturation of those structures by converting their unpacked forms to the Bell scenario à la Prop.~5 of Ref.~\cite{evans}, and then unpacking a second time.


Of course, unpacking supplemented with inflation is far more powerful than unpacking alone. Unpacking and inflation are both naturally formulated as linear programs, and hence can be easily combined into a single composite linear program to solve Causal Compatibility or Causal Optimization (over polynomials of conditional distributions).

\section{Conclusions}\label{conclusions}
The inflation technique was first proposed by~\citet{wolfe2016inflation} as a means to obtain strong causal compatibility constraints for arbitrary causal structures. Here, we have formulated inflation as a formal hierarchy of problems for assessing causal compatibility relative to correlation scenarios. We have proven the inflation hierarchy to be complete, in the sense that any distribution incompatible with a given correlation scenario will be detected as incompatible by inflation. More quantitatively, we showed that any distribution $P$ passing the $n^{th}$-order inflation test is ${O}{\left(\frac{1}{\sqrt{n}}\right)}$-close in Euclidean norm to some other distribution which \emph{can} be realized within the considered scenario. 

The inflation technique is fully applicable to any causal structure, since unpacking allows one to map any causal assessment problem (for either distributions or functions) to an equivalent assessment problem relative to a correlation scenario. The observed distribution in the original structure is mapped to probabilities pertaining to restricted sets of measurable events in the unpacked correlation scenario. Since, however, our proof of the convergence of inflation allowed for restricted sets of measurable events, the convergence theorems are still applicable when using inflation to assess compatibility relative to general causal structures. 

We have therefore shown that the inflation technique is much more than a useful machinery to derive statistical limits; it is an alternative way to define causal compatibility!

For the purpose of practical causal discovery, we envision the inflation technique being used as \emph{final refinement}. That is, inflation (and unpacking) should be employed as a postprocessing, after first filtering the set of candidate causal explanations by means of computationally-cheaper but less-sensitive algorithms. Indeed, our attitude concerning the primacy of single-district graphs reflects our implicit assumption that all the kernels of a multi-district graph will have been identified. In other words, whatever distribution is being assessed for causal compatibility via inflation, we are presuming that it has already been verified to satisfy the nested Markov property (NMP) relative to the considered graph~\cite{extraRef4,Shpitser2014NMP}. Thus, we envision testing for compatibility via inflation only \emph{after} first testing for compatibility via NMP algorithms. This is not strictly necessary, as our results here imply that the inflation technique alone can recover all the constraints implied by NMP, though we imagine it is relatively inefficient to impose NMP only indirectly through inflation.

Alternatively, inflation could be used to estimate the distances of a distribution $P$ from the sets of distributions compatible with various causal structures. We speculate that such distances could prove valuable in helping compute scores for the \emph{ranked} causal discovery problem~\cite{Magliacane2016,Evans2018ModelSelection}, though we defer further analysis to future research.
\smallskip

 
\begin{acknowledgments}
We thank T. Fritz, T.C. Fraser, A. Ac{\'i}n, and A. Pozas-Kerstjens for interesting discussions. This research was supported in part by Perimeter Institute for Theoretical Physics. Research at Perimeter Institute is supported in part by the Government of Canada through the Department of Innovation, Science and Economic Development Canada and by the Province of Ontario through the Ministry of Colleges and Universities. This work was not supported by the European Research Council. 
\end{acknowledgments}

\onecolumngrid
\bibliographystyle{apsrev4-2-wolfe}
\nocite{apsrev41Control}
\bibliography{InflationConverges}


\appendix

\appendixpage
\setcounter{equation}{0}
\renewcommand{\theequation}{A\arabic{equation}}


The purpose of this Appendix is to prove Theorem \ref{ext^definetti}.

We will first prove the result for the triangle scenario; the generalization to arbitrary correlation scenarios will be obvious. Given a distribution of $3n^2$ variables $Q(\{A^{i,j}, B^{k,l}, C^{p,q}\})$, consider its symmetrization $\tilde{Q}$, defined by
\begin{align}\label{symmetrization}
&\tilde{Q}(\{A^{ij}{=}a^{ij}, B^{kl}{=}b^{kl}, C^{pq}{=}c^{pq}\})=\frac{1}{n!^3}\Bigg(
{\smashoperator{{\sum}_{\quad\qquad\pi,\pi',\pi''\in S_n}}  Q(\{A^{ij}{=}a^{\pi(i)\pi'(j)},B^{kl}{=}b^{\pi'(k)\pi''(l)},
 C^{pq}{=}c^{\pi''(p)\pi(q)}\})}{\Bigg)}.
\end{align}
Note that $\tilde{\tilde{Q}}=\tilde{Q}$. In addition, any distribution $Q$ satisfying the symmetry condition (\ref{symmetry}) fulfills $\tilde{Q}=Q$ and any symmetrized distribution satisfies (\ref{symmetry}). 

Let $\id_{\{\hat{a}(i,j),\hat{b}(k,l),\hat{c}(p,q)\}}$ be the deterministic distribution assigning the values $\hat{a}(i,j),\hat{b}(k,l),\hat{c}(p,q)$ to the random variables $A^{i,j},B^{k,l},C^{p,q}$, for $i,j,k,l,p,q\in\{1{,}...,n\}$. Since any distribution is a convex combination of deterministic points, it follows that any distribution satisfying Eq.~(\ref{symmetry}) can be expressed as a convex combination of symmetrized distributions of the form $\tilde{\id}_{\{\hat{a}(i,j),\hat{b}(k,l),\hat{c}(p,q)\}}$. 

For clarity of notation, let us assume that the values $\{\hat{a}(i,j),\hat{b}(k,l),\hat{c}(p,q)\}$ are fixed and denote the symmetrization of $\id_{\{\hat{a}(i,j),\hat{b}(k,l),\hat{c}(p,q)\}}$ by $\tilde{P}$. Call $\tilde{P}^1$ its diagonal marginal of degree $1$, i.e., $\tilde{P}(A^{1,1},B^{1,1},C^{1,1})$. It can be verified, by symmetry, that $\tilde{P}^1$ is given by the formula:
\begin{align}
&\tilde{P}^1(a,b,c)=\frac{1}{n^3}
\smashoperator[r]{{\sum}_{i,j,k=1}^n} \delta(\hat{a}(i,j),a)\delta(\hat{b}(j,k),b)\delta(\hat{c}(k,i),c)
,
\end{align}
where $\delta(i,j)$ denotes the Kronecker delta function, i.e., $\delta(i,j)=1$ if $i=j$ or zero otherwise. Notice that $\tilde{P}^1(a,b,c)$ can be reproduced in the triangle scenario. Indeed, the latent variables are $i,j,k$, they can take values in $\{1{,}...,n\}$ and are uniformly distributed. The observed variables $a,b,c$ are deterministic functions of $(i,j)$, $(j,k)$ and $(k,i)$, respectively.

Consider now the diagonal marginal of degree $g$, ${\tilde{P}^g\equiv\tilde{P}(A^{1,1},B^{1,1},C^{1,1}{,}...,A^{g,g},B^{g,g},C^{g,g})}$. By symmetry, it is expressed as:
\begin{align}
&\tilde{P}^g(a^1,b^1,c^1{,}...,a^g,b^g,c^g)=
\frac{1}{n^3(n{-}1)^3...(n{-}g{+}1)^3}
\sum_{\bar{i},\bar{j},\bar{k}} \;
\prod_{x=1}^g\delta(\hat{a}(i^x,j^x),a^x)\delta(\hat{b}(j^x,k^x),b^x)\delta(\hat{c}(k^x,i^x),c^x),
\end{align}
where the sum is taken over all tuples $\bar{i},\bar{j},\bar{k}\in\{1{,}...,n\}^g$ with no repeated indices, i.e., such that $i^x\not=i^y$, $j^x\not=j^y$, $k^x\not=k^y$ for $x\not=y$.

Now, compare $\tilde{P}^g$ with the degree-$g$ lifting $(\tilde{P}^1)^{\otimes g}$. It is straightforward that
\begin{align}
&(\tilde{P}^1)^{\otimes g}(a^1,b^1,c^1{,}...,a^g,b^g,c^g)=\prod_{x=1}^g\tilde{P}^{1}(a^x,b^x,c^x)
=\frac{1}{n^{3k}}
\sum_{\bar{i},\bar{j},\bar{k}}\; \prod_{x=1}^g\delta(\hat{a}(i^x,j^x),a^x)\delta(\hat{b}(j^x,k^x),b^x)\delta(\hat{c}(k^x,i^x),c^x),
\end{align}
where, this time, the sum contains all possible tuples $\bar{i},\bar{j},\bar{k}\in\{1{,}...,n\}^g$. The total variation distance between the two distributions is bounded by $\frac{1}{n^3(n-1)^3...(n-g+1)^3}-\frac{1}{n^{3g}}$ times the number of tuples with non-repeated indices (namely, $n^3(n-1)^3...(n-g+1)^3$), plus $1/n^{3g}$ times the number of tuples with repeated indices (namely, $n^{3g}-n^3(n-1)^3...(n-g+1)^3$). The result is
\begin{align}
&{D}{\left(\tilde{P}^g,(\tilde{P}_1)^{\otimes g}\right)}\leq 2\left(1-\frac{n^3(n-1)^3...(n-g+1)^3}{n^{3g}}\right).
\end{align}

Finally, let $Q_n$ be any distribution satisfying Eq.~(\ref{symmetry}). Then, $Q_n=\sum_{\mu}p_\mu \tilde{P}_\mu$, where $p_\mu\geq 0$, $\sum_\mu p_\mu=1$, and $\tilde{P}_\mu$ is the result of symmetrizing $P_\mu=\id_{\{\hat{a}_\mu(i,j),\hat{b}_\mu(k,l),\hat{c}_\mu(p,q)\}}$, for some values $\{\hat{a}_\mu(i,j),\hat{b}_\mu(k,l),\hat{c}_\mu(p,q)\}$. By convexity of the total variation distance, we have that
\begin{align}
&{D}{\left(Q_n^g,\sum_\mu p_\mu (\tilde{P}_\mu^1)^{\otimes g}\right)} \nonumber
\\&\leq \sum_\mu p_\mu {D}{\left(\tilde{P}^g_\mu,(\tilde{P}_\mu^1)^{\otimes g}\right)}\nonumber\\
&\leq 2\left(1-\frac{n^3(n-1)^3...(n-g+1)^3}{n^{3g}}\right)\nonumber\\
&=2\left(1-\left[\frac{n^g-n^{g-1}(1{+}2{+}...{+}g{-}1)+O(n^{g-2})}{n^{g}}\right]^3\right)\nonumber\\
&=2\left(1-\left[1-\frac{O(g^2)}{n}\right]^3\right)\nonumber\\
&={O}{\left(\frac{3g^2}{n}\right)}.
\end{align}

Extending this result to general correlation scenarios is straightforward, so we will just sketch the proof. First, the action of the corresponding symmetrization over a deterministic distribution equals a distribution $\tilde{P}$ whose $1$-marginal $\tilde{P}^1(a_1{,}...,a_m)$ is a uniform mixture over the tuple of indices $\bar{i}$ of deterministic distributions of the form $\prod_{x=1}^m\delta(a_x,a_x(\bar{i}^{L_x}))$. Again, we remind the reader that $L_x\subset\{1,...,L\}$ denotes the indices of the hidden variables on which $A_x$ depends. It thus follows that $\tilde{P}^1$ is realizable within the correlation scenario. The diagonal marginal $\tilde{P}_g$ is also a uniform mixture of deterministic distributions of a similar type, but where no repeated indices are allowed between the different blocks of variables. The statistical difference between $\tilde{P}^g$ and $(\tilde{P}^1)^{\otimes g}$ is thus bounded by
\be
2\left(1-\frac{n^L(n-1)^L...(n-g+1)^L}{n^{gL}}\right)={O}{\left(\frac{Lg^2}{n}\right)}.
\ee
As before, the general result follows from the convexity of the total variation distance.

\end{document}